\documentclass[aps,prc,reprint,showpacs,groupedaddress,onecolumn]{revtex4-1}

\usepackage{graphicx,color} 
\usepackage{dcolumn}  
\usepackage{bm}       
\usepackage{amsmath}
\usepackage{xcolor}

\begin{document}

\newcommand {\nc} {\newcommand}

\newcommand{\vv}[1]{{$\bf {#1}$}}
\newcommand{\ul}[1]{\underline{#1}}
\newcommand{\vvm}[1]{{\bf {#1}}}
\def\btau{\mbox{\boldmath$\tau$}}

\nc {\IR} [1]{\textcolor{red}{#1}}
\nc {\IB} [1]{\textcolor{blue}{#1}}
\nc {\IP} [1]{\textcolor{violet}{#1}}
\nc {\IG} [1]{\textcolor{olive}{#1}}
\nc {\IT} [1]{\textcolor{teal}{#1}}

\title{Few-body universality in the deuteron-alpha system}

\author{Jin Lei$^{(a)}$}
\email{jinl@ohio.edu}
\author{L.~Hlophe$^{(b)}$}
\email{hlophe@nscl.msu.edu}
\author{Ch.~Elster$^{(a)}$}
\author{A. Nogga$^{(c)}$}
\author{F.M.~Nunes$^{(b)}$}
\author{D.R.~Phillips$^{(a,d,e)}$}

\affiliation{
(a)Institute of Nuclear and Particle Physics,  and
Department of Physics and Astronomy,  Ohio University, Athens, OH 45701,
USA \\
(b) National Superconducting Cyclotron Laboratory and Department of Physics and Astronomy, Michigan State University, East Lansing, MI 48824, USA \\
(c) IAS-4, IKP-3, JHCP, and JARA-HPC,  Forschungszentrum J\"ulich, D-52428
J\"ulich, Germany\\
(d) Institut f\"ur Kernphysik, Technische Universit\"at Darmstadt, 64289 Darmstadt, Germany\\
(e) ExtreMe Matter Institute EMMI, GSI Helmholtzzentrum f{\"u}r Schwerionenforschung GmbH, 64291 Darmstadt, Germany}

\date{\today}

\begin{abstract}
We treat ${}^6$Li as an effective three-body ($n$-$p$-$\alpha$) 
system and compute the $d$-$\alpha$ $S-$wave scattering length and 
three-body separation energy of ${}^6$Li for a wide variety of nucleon-nucleon and $\alpha$-nucleon 
potentials which have the same (or nearly the same) phase shifts. 
The Coulomb interaction in the $p$-$\alpha$
subsystem is omitted. The results of all calculations lie on a 
one-parameter curve in the plane defined by the $d$-$\alpha$ $S-$wave 
scattering length and the amount by which ${}^6$Li is bound with respect 
to the $n$-$p$-$\alpha$ threshold. We argue that these aspects of the 
$n$-$p$-$\alpha$ system can be understood using few-body universality and 
that ${}^6$Li can thus usefully be thought of as a two-nucleon halo nucleus.
\end{abstract}


\maketitle

\paragraph*{Introduction}

Few-body universality is a powerful tool to analyze the low-energy properties of quantum mechanical
systems that are weakly bound~\cite{Braaten:2004rn,Konig:2016utl}. 
Applications of few-body universality range from atomic and molecular
physics, e.g., atomic species near a Feshbach resonance~\cite{Ferlaino:2009zz} or dimers and trimers of $^4$He atoms~\cite{Kunitski:2015qth}, to nuclear physics, 
e.g., few-nucleon systems~\cite{Bedaque:2002mn} and halo nuclei~\cite{Hammer:2017tjm}, to hadronic physics, e.g., the $X(3872)$ and other ``exotic" mesons near two-meson thresholds.
All these systems have in common that their two-body separation energy is small enough that the wave function of the
effective low-energy degrees of freedom (e.g., atoms, nucleons, $D$ and $\bar{D}$-mesons) has much of its support
in a region outside the interaction potential, i.e., in the tunneling regime. The properties of the two-body systems
are then, to a first approximation, independent of details of the potential, and are correlated solely 
with the separation energy. The qualitative picture of two-body universality laid out in this paragraph can be
systematically organized in terms of an effective field theory (EFT) expansion in powers of $R$, the range of the
two-body potential, times $\gamma=\sqrt{2 \mu E_2}$, the binding momentum of the two-body bound state (with $E_2$ the two-body
separation energy and $\mu$ the two-body reduced mass) since $\gamma$ determines the exponential fall-off of the two-body wave function outside the potential. 

Many of these systems also exhibit three-body bound states. However, the three-body 
separation energy $E_3$ is not solely determined by the two-body separation energy, 
although it does depend on it. At leading order (LO) in the $\gamma R$ expansion, 
one three-body observable must be used to fix a ``three-body parameter". 
All other properties of the three-body system are then determined by the three-body observable chosen (e.g., the separation energy $E_3$) and $E_2$~\cite{Bedaque:1998km,Bedaque:1998kg,Bedaque:1999ve}. 
It is important to note---especially in the context
of our calculation presented below---that the
three-body parameter need not arise from ``intrinsic" three-body forces. It may, instead, in part or in whole, reflect off-shell
properties of two-body forces that are not observable in the two-body system, 
and  first have experimental consequences in the three-body system~\cite{Naidon:2014oma,Polyzou:90}. 
 If $E_2$ is small compared to $E_3$ and $\sqrt{2 \nu E_3} R$ (with $\nu$ the 2+1 reduced mass) is also small, 
then there is the possibility to observe a sequence of three-body bound states, which are related to one another
by a scaling transformation, as predicted by Efimov~\cite{Efimov:1970zz,Efimov:1979zz}. But, even in systems where the conditions for the emergence of bound excited Efimov states are not met, universality still  
connects disparate three-body systems to one another and provides insights that aid in organizing their phenomenology~\cite{Hammer:2010kp,Naidon:2016dpf}.

For example, one important consequence of universality in the three-body system is that $E_3$ is correlated with the scattering length
of the third particle from the two-body bound state.  This correlation persists to much smaller 2+1 scattering 
lengths $a_{21}$ than does the correlation obtained by considering the 
three-body system to be weakly bound with respect to the 2+1 threshold,  $E_3=\frac{1}{2 \nu a_{21}^2} + E_2$.
  In the three-nucleon system, the $E_3$-$a_{21}$ correlation---which in this case is with the scattering length in the total-spin-$1/2$ channel, where the three-body bound
state, the triton, resides---was first demonstrated
by Phillips~\cite{Phillips:1968zze} and is known as ``the Phillips line". This ``Phillips line" still emerges for nucleon-nucleon ($NN$) potentials that are fitted much more accurately to data than were those originally examined by Phillips~\cite{Witala:2003en}.
Efimov~\cite{Efimov:1998fb} demonstrated that such a correlation
is a consequence of the shallow binding of the two-body system, and it has been computed at LO
and next-to-leading order (NLO) in the EFT that encodes universality in the 
three-nucleon system~\cite{Bedaque:1999ve,Hammer:2001gh}. 

In this paper we show that a similar, universal, correlation occurs between the three-body separation energy
of $^6$Li  and the $d$-$\alpha$ $S-$wave scattering length $a_{d\alpha }$. 
We do this by modeling the $d$-$\alpha$ system as an effective three-body problem, 
in which the neutron, proton, and $\alpha$-particle are viewed as basic 
degrees of freedom that interact via pairwise forces.
This is justified because the first excited state of the $\alpha$ particle is $\approx 20$ MeV above its ground state and
the $\alpha$ particle is compact with respect to $^6$Li. Our ansatz follows a large body of work treating $^6$Li as a
three-body problem, see e.g.~\cite{Eskandarian:1992zz,Lehman:1982zz,Lehman:1982zz,Lehman:1978zz,Schellingerhout:1993ku}.

We note that there is also a study of the implications of universality for
${}^6$Li as a six-body system. In Ref.~\cite{Stetcu:2006ey} Stetcu, Barrett, 
and van Kolck constructed an EFT for the No-Core Shell Model and 
determined the leading-order $NN$ and three-nucleon forces in the EFT by 
demanding that the experimental binding energies of the deuteron, triton, and $\alpha$-particle
are exactly reproduced. Their six-body calculation then had ${}^6$Li 
unbound with respect to the $d$-$\alpha$ threshold; $a_{d\alpha }$  thus could not be computed. 
In contrast, our three-body model of ${}^6$Li avoids the need to compute 
the emergent low-energy scales in ${}^5$He and ${}^6$Li {\it ab initio} from $NN$ and three-nucleon
forces. Instead, it takes those scales as input and elucidates their 
consequences for the low-energy dynamics of the $d$-$\alpha$ system.

For the purpose of this work we ignore the Coulomb effects
between the $\alpha$ particle and the proton. 
The $\alpha$ interacts with the nucleons predominantly in $P$-waves, 
while the neutron and proton interaction is mainly $S$-wave. 
The resulting three-body system thus has different dynamics to the 
three-nucleon case described above, since it contains two $P$-wave attractive 
interactions and only one $S$-wave one.

\paragraph*{Framework} 
We take the neutron-proton ($np$) force in the $^3S_1$-$^3D_1$ channel, and the $\alpha N$ force
in the $P_{3/2}$, $P_{1/2}$, and $S_{1/2}$ partial waves. The three-body separation energy of $^6$Li is
obtained by solving bound state Faddeev equations with separable representations of these
forces as outlined in Ref.~\cite{Hlophe:2017bkd}. (The ``three-body separation energy" of $^6$Li is the amount
by which it is bound compared to the $n$-$p$-$\alpha$ threshold, and thus is equal to 
its  $d$-$\alpha$ separation energy plus the $n$-$p$ separation energy of deuteron.)
The work of Ref.~\cite{Hlophe:2017bkd} showed that in this system the 
solution of the Faddeev equations with separable forces is numerically 
indistinguishable from the solution with non-separable forces provided
the separable basis is appropriately chosen.

For $d$-$\alpha$ scattering, we solve the momentum space Faddeev-AGS equations~\cite{ags},
\begin{equation}
\label{eq:ags}
U_{ij}(E)=\bar{\delta}_{ij}G_0^{-1} (E) + \sum_{k=1}^3 \bar{\delta}_{ik}
t_{k}(E)G_0(E)U_{kj}(E),
\end{equation} 
with $\bar{\delta}_{ij}=1-\delta_{ij}$, and  $G_0(E)=(E+i0 -H_0)^{-1}$ being
the free resolvent at the available energy $E$. The free three-particle
Hamiltonian is denoted by $H_0$, while $t_k = v_k +v_k G_0(E) t_k$ is the two-body transition matrix. Here the index $k$ stands for the channel corresponding to
the configuration where the particle $k$ is the spectator and the remaining two form the pair
($ij$). Since here we are working with three distinguishable particles, cyclic permutations of
($ijk$) leads to the three required transition operators in Eq.~(\ref{eq:ags}). Since we are
interested only in very low energy scattering, we do not have to treat breakup singularities,
and the numerical solution of Eq.~(\ref{eq:ags}) is straightforward. As in the bound state
calculation~\cite{Hlophe:2017bkd} we employ the separable representation of the interactions in the two-body
subsystems, which was shown to lead to numerically the same observables as a solution with non-separable
forces for continuum~\cite{Cornelius:1990zz}. In addition, we employ the same model space in the scattering
calculation as is used to calculate the three-body separation energy of $^6$Li; this is sufficient when
studying the low energy parameters in the $d$-$\alpha$ channel with $J^\pi=1^+$ and total isospin $T=0$.

In order to investigate if there is  a correlation between the three-body
separation energy of $^6$Li and the $d$-$\alpha$ $S-$wave scattering length, one 
needs to solve for these quantities using different sets of potentials 
which describe the low-energy behavior in the subsystems with the same 
quality, i.e., potentials that are phase shift equivalent. In the case of the 
$np$ interaction this is relatively easy to achieve, since all modern
$NN$ interactions are fitted to describe the deuteron binding energy, the $np$ low
energy parameters (scattering length and effective range) and phase shifts in the energy range we are
considering. The situation is quite different in the case of effective 
$\alpha N$ interactions. There have been several efforts to construct effective 
$\alpha N$ interactions of varying degrees of sophistication (e.g. \cite{kanada:79, garrido:97,Bang:1983xpz,Bang:1979ihm}).
However, the condition of phase shift equivalence was imposed rather 
loosely compared to the $NN$ subsystem. Thus we need to consider a different approach to construct
phase shift equivalent $\alpha N$ potentials. Following the suggestion of
Refs.~\cite{Afnan:1973znm,Thomas:1975qfg} we employ a unitary transformation (UT) of the $\alpha N$
Hamiltonian $H_{2b}=h_0+v$ with $h_0$ being the two body kinetic energy operator and $v$ the effective two-body
interaction. Following \cite{Afnan:1973znm,Thomas:1975qfg} we define a transformed Hamiltonian
\begin{equation}
\label{eq:ut}
\tilde{H}_{2b} = U H_{2b} U^\dagger = h_0 + \tilde{v},
\end{equation}
where $\tilde{v}$ is the transformed potential keeping the phase shifts unchanged. The operator
for the UT is defined as
\begin{equation}
\label{eq:Uoperator}
U=1-2|h\rangle \langle h| .
\end{equation}
Following Ref.~\cite{Haftel:1971er} we
choose for $ | h\rangle$
\begin{equation}
\label{eq:h_func}
\langle r Y_l^m| h\rangle = N r^l {\rm e}^{-c r} (1-b r),
\end{equation}
where $N$ is evaluated through the normalization condition $\langle h | h \rangle =1$
for each partial wave. 
In our calculations, we only consider the UT on $P-$waves. 
We include the factor of $r^l$ in accord with Ref.~\cite{Haftel:1971er}, and pick $b=1$~fm$^{-1}$ for simplicity.
We vary the parameter  $c$, thereby changing the range of the transformation.  If the starting potential $v$ is separable and of
rank-1, the transformed potential $\tilde{v}$ will be of rank-3~\cite{Afnan:1973znm}. In the case
of an arbitrary local or nonlocal $v$, the transformed potential will have to be numerically
calculated, leading to a nonlocal potential $\tilde{v}$. 

\paragraph*{Results}
To study a possible correlation between the three-body separation energy of $^6$Li and the
corresponding $S-$wave scattering length in the $d$-$\alpha$ channel, we start by using very simple,
rank-1 separable interactions in the two-body subsystems. The form factors of the
separable interactions are of Yukawa type, and the parameters are fitted to reproduce the
deuteron binding energy and $np$ low-energy scattering parameters in the case of the $np$ interaction, and
the $\alpha N$ $S-$ and $P-$wave phase shifts up to 10~MeV in the case of the $\alpha N$ interaction.
Specifically, for the $\alpha N$ interaction we
employ model A from Ref.~\cite{Eskandarian:1992zz} and for the $np$
interaction we choose the parameters from that work that lead to a deuteron $D-$state probability of 4\%. 

\begin{figure}[tb]
\begin{center}
\includegraphics[scale=.55]{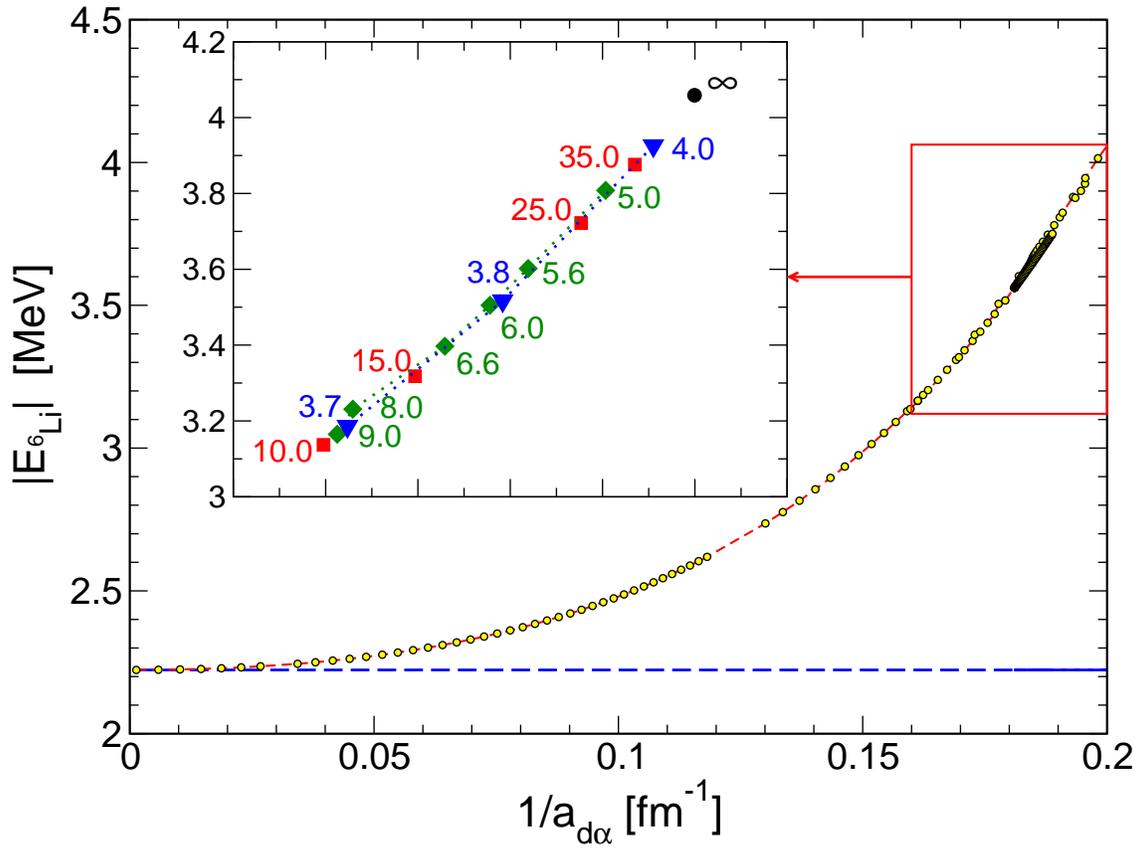}
\caption{\label{fig:fig1} (Color online) The absolute value of the three body separation energy of
$^6$Li as function of the inverse of the
$d$-$\alpha$ $S$-wave scattering length $a_{d\alpha}$ for phase shift equivalent interactions obtained by
unitarily transform the interactions in the n$\alpha$ $P_{3/2}$ and $P_{1/2}$ channels.
The insert magnifies the marked rectangle and indicates the value $c$ of the exponent in the 
transformation of Eq.~(\ref{eq:h_func}). The dashed horizontal line indicates the deuteron
breakup threshold.
}
\end{center}
\end{figure}

We then apply the UT of Eqs.~(\ref{eq:Uoperator}) and~(\ref{eq:h_func}) to
the $P-$waves of the $\alpha N$ interaction and reduce the parameter $c$ in Eq.~(\ref{eq:h_func}),
starting from a value $c$~=~35~fm$^{-1}$ until we reach values at which $^6$Li is no longer bound. 
The result of these calculations is summarized in
Fig.~\ref{fig:fig1}, which shows the dependence of the 
three-body separation energy
of  $^6$Li as a function of the inverse $S-$wave scattering length $a_{d\alpha}$.
(Almost exactly the same correlation of inverse scattering length and three-body separation energy is obtained if the UT is only employed in the $P_{3/2}$ channel, and a very similar result is obtained if only the $P_{1/2}$ $\alpha N$ partial wave is unitarily transformed.)
The insert magnifies the regime when $c$ varies from 35~fm$^{-1}$ to 4~fm$^{-1}$, and also
shows the calculation using the unmodified $\alpha N$ interaction as a solid circle (labeled by $\infty$). First, a
decrease in $c$ from 35~fm$^{-1}$ to 10~fm$^{-1}$ leads to a decrease in the $^6$Li separation
energy together with an increase in the scattering length, forming a line along which
the loci of separation energy versus inverse scattering length sit (red solid squares). 
When $c$ is further decreased, this trend reverses, with the loci now following the previous line, but in the
opposite direction---as indicated by the green diamonds in the inset of Fig.~\ref{fig:fig1}. This
phenomenon of directional reversal on the correlation line has also been observed in Ref.~\cite{Thomas:1975qfg}, where the UT
was applied to $NN$ potentials in the three-nucleon problem.
Once the value of
$c$  drops below 4~fm$^{-1}$, the separation energy decreases uniformly as a function of the inverse
scattering length until the deuteron breakup threshold is reached at $c$~=~3.9~fm$^{-1}$. At this point $^6$Li becomes unbound, and $1/a_{d\alpha} \rightarrow 0$. Figure~\ref{fig:fig1} shows that all calculations determine a single parametric curve. 

\begin{figure}[tb]
\begin{center}
\includegraphics[scale=.55]{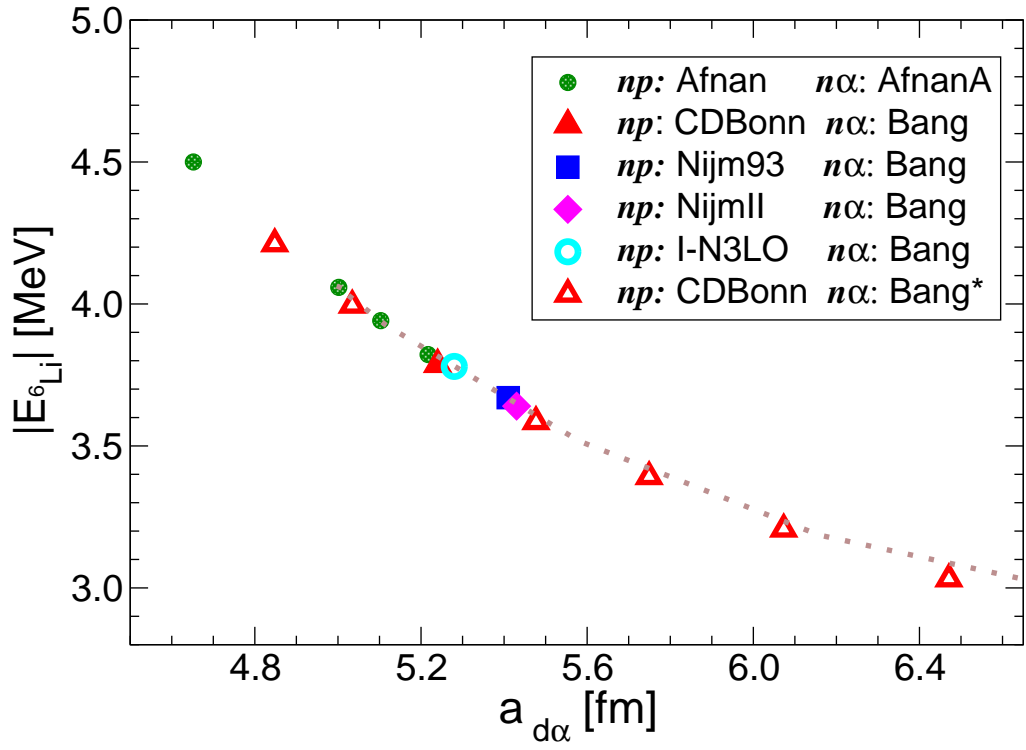}
\caption{\label{fig:fig2} (Color online) The absolute value of the three body separation energy 
of $^6$Li
as a function of $d$-$\alpha$ $S-$wave scattering length $a_{d\alpha}$ calculated with a variety of interactions,
as described in the text and indicated by the legend. The faint dotted line picks up points
from Fig.~\ref{fig:fig1} and is intended to guide the eye.}
\end{center}
\end{figure}

The large variation of the parameter $c$ in the UT of the $\alpha N$
interaction in the $P-$wave may appear somewhat artificial. Thus as the next step we consider 
``realistic" interactions in the two-body sub-system. For the $\alpha N$ interaction we choose
the Bang interaction~\cite{Bang:1983xpz}, where we set the strength parameter of the
central Woods-Saxon term to -44~MeV as in Ref.~\cite{Hlophe:2017bkd}, 
while for the $np$ interaction we employ the CD-Bonn potential~\cite{Machleidt:2000ge}.
This $\alpha N$ interaction generates a Pauli-forbidden $S$-wave $\alpha N$ bound state, 
which we remove from the two-body spectrum using the methods described in Ref.~\cite{Hlophe:2017bkd}. 
Omitting the Coulomb interaction we then obtain a ${}^6$Li three-body separation energy
of $-3.78$~MeV and a scattering length of 5.29~fm, indicated in Fig.~\ref{fig:fig2} 
as a solid red upward triangle. As a guide to the eye a
subset of the points from Fig.~\ref{fig:fig1} is also displayed in Fig.~\ref{fig:fig2} as a faint dotted line; we see
that this calculation based on ``realistic" interactions falls almost on top of the line determined
previously by the rank-1 separable interactions. This indicates that off-shell/high-momentum details of the 
two-body forces do not influence the low energy behavior
of the $d$-$\alpha$ system---except to the extent that a particular force's high-momentum behavior
determines the particular point on the correlation line at which it resides. To check if this is indeed the case, we 
employ a series of $np$
interactions which have quite different off-shell/high-momentum behavior but are
all fitted to the deuteron binding energy and the $^3S_1$-$^3D_1$ phase shift with high
precision. The calculation based on the Nijmegen-93 potential~\cite{Stoks:1994wp} is
indicated by the blue solid square, the Nijmegen-II potential~\cite{Stoks:1994wp} by the magenta solid
diamond, and the Idaho-N4LO potential~\cite{Entem:2003ft} by the open cyan circle. Though the
realistic $NN$ interactions are located very close to each other in Fig.~\ref{fig:fig2}, they 
all fall on the line established by the previous calculations shown in Fig.~\ref{fig:fig1}. In
addition to the modern $NN$ interactions we also include as filled green circles the rank-1
$np$ interaction from Ref.~\cite{Eskandarian:1992zz} in which the deuteron $D$-state probability is
varied for the $np$ interaction. 

In order to further explore this behavior for more sophisticated potentials we also modify 
the strength of the Woods-Saxon potential in the central part of the Bang
$\alpha N$ interaction from -42~MeV to -45~MeV; this preserves the general characteristics of the $\alpha N$ system, 
i.e., leaves it unbound, but causes the agreement with the $\alpha N$ phase shifts to deteriorate and the P$_{3/2}$ 
resonance position to move. 
Keeping the $np$ interaction fixed while making this change yields results for the three-body system that 
are represented in Fig.~\ref{fig:fig2} by the red open upward triangles. They are 
consistent with the line established earlier. This is a non-phase-equivalent variation of the Bang interaction, so 
it is somewhat surprising that the $E_{^6{\rm Li}}$-$a_{d\alpha}$ curve is unaffected. In contrast, changing the strength of the $NN$ interaction, so altering the deuteron binding energy, yields a $E_{^6{\rm Li}}$-$a_{d\alpha}$ curve whose linear portion has a different slope (not shown). The correlation seems to be more sensitive to the on-shell $NN$ input than it is to the on-shell $\alpha N$ input. 

\paragraph*{Interpretation and Implications}
We compute the universal correlation between $a_{d\alpha}$ and $E_{^6{\rm Li}}$ by evaluating both quantities using several
$np$ potentials that have different high-momentum/off-shell behavior, but almost
the same $np$ phase shifts, together with a continuous family of
 $\alpha N$ potentials that have different high-momentum/off-shell
behavior but exactly the same $\alpha N$ phase shifts.
Arbitrary combinations of these two-body potentials yields results for the three-body observables that lie on a single
curve in the $a_{d\alpha}$- $E_{^6{\rm Li}}$ plane.

The $a_{d\alpha}$-$E_{{}^6{\rm Li}}$ correlation displayed here is certainly related to the well-known ``Phillips line" 
of the neutron-deuteron system: it is not surprising that $NN$ interactions with different off-shell behavior
produce points along a curve in the  $a_{d\alpha}$-$E_{{}^6{\rm Li}}$ plane. 
The novel feature of the $n$-$p$-$\alpha$ system is that varying the off-shell properties
of the $P$-wave nucleon-$\alpha$ potential also produces points on the same curve. 
This kind of correlation is typical
of weakly-bound systems and is a consequence of few-body universality. It is in accord with analyses of
$^6$He that show universal correlations are expected for weakly bound, three-body systems where the same
angular-momentum-structure
of two-body potentials occurs as in $^6$Li~\cite{Rotureau:2012yu,Ji:2014wta}. 

The existence of an $a_{d\alpha}$-$E_{{}^6{\rm Li}}$ correlation thus suggests that
$^6$Li can be thought of as a ``deuteron halo".
Indeed, the 
experimental  $d$-$\alpha$ separation energy of $^6$Li (1.47 MeV)~\cite{AjzenbergSelove:1988ec} is  
comparable to the deuteron binding energy ($B_d=2.22$ MeV), and certainly
much smaller than the energy associated with $\alpha$-particle excitation. Recent work on infra-red 
extrapolations of the ${}^6$Li binding energy in {\it ab initio} No-core Shell Model  calculations using sophisticated $NN$ and three-nucleon forces
also show a typical momentum that
is much smaller than that of the $\alpha$ particle, supporting its identification as a halo nucleus~\cite{Forssen:2017wei}. 

The portion of the curve at very large $a_{d \alpha}$, i.e., very small deuteron separation energy, is well described by an effective-range expansion in the $d \alpha$ system.
However, such a two-body description is only valid when $|E_{{}^6 {\rm Li}}| - B_d < B_d$, i.e. the deuteron separation energy of ${}^6$Li is significantly less than the deuteron binding energy. 
When ${}^6$Li is more bound the $a_{d\alpha}$-$E_{{}^6{\rm Li}}$ correlation is linear, with a slope that depends on low-energy $NN$ observables. In this domain
changes of the $NN$ interaction that alter the $NN$ phase shifts and the deuteron binding energy yield a different relation between $a_{d\alpha}$ and $E_{{}^6{\rm Li}}$. We conclude that, at least for realistic ${}^6$Li binding, the connection 
between $a_{d\alpha}$ and $E_{{}^6{\rm Li}}$ is a consequence of universality in the three-body $n$-$p$-$\alpha$ system, and cannot be understood using a low-order effective-range expansion for the $d$-$\alpha$ system. 

As is well known from three-nucleon systems \cite{Bedaque:1999ve}, such a strict correlation suggests that 
one three-body force can absorb the dependence on the unitary transformation at leading order in the $\gamma R$ expansion.
We caution that here we have only examined the existence of such a correlation in the $\alpha$-$n$-$p$ channel with 
total angular momentum $J=1$, positive parity, and total isospin $T=0$. But, following the example of the three-nucleon case, 
we anticipate that other low-energy $d$-$\alpha$ observables---not just $a_{d \alpha}$---are
correlated with the three-body separation energy. If that's the case then $d$-$\alpha$  scattering should be accurately predicted
starting from $\alpha$-nucleon and $np$ interactions as long as the three-body separation energy is reproduced.

In Ref.~\cite{Ryberg:2017di} Ryberg {\it et al.} performed an EFT calculation of the $\alpha n n$ system and
argued that, for the $^{6}$He channel where $J=0$ and $T=1$, there were at least two three-body force structures if both the 
$P_{3/2}$ and $P_{1/2}$ channels were included in the $\alpha N$ interaction. In contrast, 
we found that the $a_{d\alpha}$-$E_{{}^6{\rm Li}}$ correlation is very similar regardless of whether only $P_{3/2}$, only $P_{1/2}$, or both $\alpha N$ 
channels are unitarily transformed. Thus we have no indication that a second three-body 
force structure contributes to low-energy $\alpha d$ observables at leading order in the $\gamma R$ expansion, even if both $P$-wave $\alpha N$ channels are included non-perturbatively
in the three-body calculation. The extent to which other observables are correlated with the $^{6}$Li binding energy is an interesting topic for future work, as is the identification of 
the leading three-body force in all of the ${}^6$Li three-body channels~\cite{Griesshammer:2005ga}. 
 
As mentioned before, we have not included the Coulomb repulsion between the $\alpha$ particle and the proton in this 
analysis. It seems reasonable to expect that the halo nature of the ${}^6$Li system unveiled in this 
study will still be present once Coulomb effects are included (cf. Ref.~\cite{Hammer:2008ra} for a study of this issue in a two-body model).
In Ref.~\cite{Hlophe:2017bkd} a subset of the authors computed the amount 
by which that force reduces the three-body separation energy of $^6$Li, but those results 
 were only for the ${}^6$Li bound state.
Once we have the ability to include the Coulomb force when solving the scattering 
Faddeev-AGS equations with separable interactions, it will be worthwhile revisit 
the calculations present here and assess the impact of the repulsive $\alpha p$ 
electrostatic interaction on the universal correlations in the ${}^6$Li system.

\begin{acknowledgments}
We thank Hans-Werner Hammer for useful discussions and for comments on the manuscript.
This work was performed in part under the
auspices of the National Science Foundation under contract NSF-PHY-1520972
with Ohio University and NSF-PHY-1520929 with Michigan State University, 
of the U.~S.  Department of Energy under contract
No. DE-FG02-93ER40756 with Ohio University, and of DFG and NSFC through funds provided to the
Sino-German CRC 110 ``Symmetries and the Emergence of Structure in QCD" (NSFC
Grant No.~11621131001, DFG Grant No.~TRR110), and by the 
 ExtreMe Matter Institute EMMI
at the GSI Helmholtzzentrum f\"ur Schwerionenphysik, Darmstadt, Germany.
The numerical computations benefited from resources of the National
Energy Research Scientific Computing Center, a DOE
Office of Science User Facility supported by the Office of Science of
the U.S. Department of Energy under contract
No. DE-AC02-05CH11231. 
\end{acknowledgments}

\bibliography{reactions,halos,hwh}

\newpage

\end{document}